# IOTSim: a Cloud based Simulator for Analysing IoT Applications

**Xuezhi Zeng[1],** [2]Saurabh Kumar Garg, [1]Peter Strazdins, [3]Prem Jayaraman,
[3]Dimitrios Georgakopoulos and [4]Rajiv Ranjan

[1]Australian National University, Australia
{xuezhi.zeng, peter.strazdins}@anu.edu.au

[2]University of Tasmania, Australia
saurabh.garg@utas.edu.au

[3]RMIT, Melbourne, Australia
{prem.jayaraman, dimitrios.georgakopoulos}@rmit.edu.au

[4]Newcastle University, UK
raj.ranjan@ncl.ac.uk

SUMMARY

A disruptive technology that is influencing not only computing paradigm but every other business is the rise of big data. Internet of Things (IoT) applications are considered to be a major source of big data. Such IoT applications are in general supported through clouds where data is stored and processed by big data processing systems. In order to improve the efficiency of cloud infrastructure so that they can efficiently support IoT big data applications, it is important to understand how these applications and the corresponding big data processing systems will perform in cloud computing environments. However, given the scalability and complex requirements of big data processing systems, an empirical evaluation on actual cloud infrastructure can hinder the development of timely and cost effective IoT solutions. Therefore, a simulator supporting IoT applications in cloud environment is highly demanded, but such work is still in its infancy. To fill this gap, we have designed and implemented IOTSim which supports and enables simulation of IoT big data processing using MapReduce model in cloud computing environment. A real case study validates the efficacy of the simulator.

1. INTRODUCTION

   According to a study by IBM, we are creating 2.5 quintillion (2.5×1018) bytes of data every day as of 2012 through different sensing devices[1, 2]. IDC (International Data Corporation) predicts that the digital universe is set to explode to an unimaginable 8 Zettabytes by 2015. It can be declared that we are in the era of "Big Data" which is accelerated by the Internet of Things (IoT) [3]. Such "Data Explosions" have led to the next grand challenge in computing known as the 'Big Data' problem [4-7], which is defined as the practice of collecting and analysing structured and unstructured data sets flowing at a volume and velocity that is too large and too fast to store, process, and interpret manually or using traditional data management applications.
   IoT applications require unique and advanced technologies to efficiently process data which are relatively large (i.e. from terabytes to exabyte) and complex (i.e. from social media to sensor data) within tolerable elapsed time. One of well-known and established technologies is distributed parallel computing (e.g. Apache Hadoop and Apache Storm) which can support IoT applications for processing batch and streaming data across parallel clusters of cloud resources[8]. For batch-oriented big data processing, the MapReduce model is a predominant programming paradigm. MapReduce is an associated implementation for processing and generating large data sets with a parallel, distributed algorithm on a cluster[9]. Typical implementations of the MapReduce model include Disco[10], Mars[11], Phoenix[12], Hadoop[13] and Google's implementation[14]. Among them, Hadoop, which is inherently designed for batch and high throughput processing jobs, has proven itself as the de facto solution to big data processing.

Recently there have been increasing IoT applications that are adapted to batch and stream computing model. However, due to their intrinsic nature, IoT applications require lots of IT resources if users want fast analysis of their large datasets. Thousands of CPUs, hundreds of terabytes of storages and very high speed interconnections are demanded. Because of this, evaluation and analysis on IoT applications in a real computing environment such as public clouds can be a challenge for several reasons:

- It is not cost-effective to procure or rent a large scale datacentre resource pool that will accurately reflect realistic application deployment and let practitioners experiment with dynamic hardware resource and big data processing framework configurations, and changing data volume, velocity, and variety
- Frequently changing experiment configurations in a large-scale real test bed involves lot of manual configuration, making the performance analysis itself time-consuming. As a result, the reproduction of results becomes extremely difficult
- The real experiments on such large-scale distributed platform are sometimes impossible due to multiple test runs in different conditions
- It is almost impractical to set up a very large cluster consisting hundreds or thousands of nodes to test the scalability of the system

An obvious solution to the aforementioned problems is to use a simulator supporting IoT application processing. A simulator not only allows us to measure scalability of computing resources for IoT applications efficiently, but also enables to determine the effects of various independent variables (i.e., datacentre configuration, virtual machine (VM) configuration, VM number, job configuration, MR combination) on different dependent variables (i.e., average execution time, maximum execution time, minimum execution time, make span, VM computation cost, network cost) which will be detailed in Section 5.2 and Section 5.3 respectively. Thus, IoT simulator will be a very useful tool to facilitate both researchers and commercial entities equally to analyse, test and design IoT applications with far less cost and time.

As the Cloudsim simulation software is the best choice to simulate cloud computing resources [15], we have designed and implemented a simulator called IOTSim on top of Cloudsim, where we can simulate the behaviour of IoT applications utilizing MapReduce framework to process the big data generated from different sensing devices. The key contributions of IOTSim lie in extending Cloudsim with 1) IoT application model support and 2) enabling processing of IoT data using big data system (i.e., MapReduce) in Cloud Computing environment. The proposed simulator also allows modelling and simulation of network usage between storage and processing virtual machines, and between individual VM.

The rest of this paper is organized as follows: Section 2 presents the general IoT architecture with its definition described and discusses the requirements for modelling IoT-based applications within a simulator. Section 3 conducts an extensive literature review of simulators in cloud computing environment and those simulators that specifically targets the MapReduce model. Section 4 details the design and implementation of the proposed IOTSim simulation framework. In Section 5, simulation results to show the efficacy of the proposed simulation tool are discussed. Section 6 concludes the paper and points out some future work.

2. REQUIREMENT FOR MODELLING IoT-BASED APPLICATIONS

## 2.1 Generic IoT Architectutre

The Internet of Things (IoT) is a network of networks, in which objects, animals or people are provided with unique identifiers and the ability to transfer data over a network without requiring human-to-human or human-to-computer interaction[16]. The IoT allows people and things to be connected anytime, anyplace, with anything and anyone through the information and communications infrastructure to provide value-added services[17]. The IoT has evolved from the convergence of wireless technologies, micro-electromechanical systems (MEMS) and the Internet. In a general way, IoT is formed by three layers[18, 19].

- Perception layer: which is the bottom layer whose function is to gather and transform data to readable digital signals with RFID, sensors, etc. All the data collection and data sensing part is done on this layer[20].
- Network layer: a middle layer which collects the data perceived by the perception layer and sends digital signals to corresponding platforms via network. This layer may only include a gateway, having one interface connected to the sensor network and another to the Internet.
- Application layer: is on the top layer, which performs the final presentation of data. Application layer receives information from the lower layer and provides global management of the application presenting that information. According to the needs of user, Application layer presents the data in the form of: smart city, healthcare, video surveillance and other many kinds of applications[21].

According to Gartner, there will be nearly 50 to 100 billion devices and sensors on the Internet of Things by 2020[17]. Once all these devices and sensors are connected with each other, IoT enables more and more new and innovation applications that support our basic needs, economies, environment and health. Such enormous number of devices connected to internet provides many kinds of services and generate Big Data [7] that needs to be processed and analysed for knowledge extraction. In order to support these IoT applications, a reliable, elastic and agile platform is essential. Cloud computing is one of the enabling platforms to support IoT applications.

Cloud computing[22] is a model for on-demand access to a shared pool of configurable resources (e.g. compute, networks, servers, storage, applications, services, and software) that can be easily provisioned by three commonly deployed cloud service models namely Infrastructure as a Service (IaaS), Platform as a Service (PaaS), Software as a Service (SaaS). For example, IaaS is able to support the perception layer by developing custom gateway interfaces to support IoT devices or sensors. Consumers can set up arbitrary services and manage the devices or sensors via cloud resource access control. PaaS can provide a platform from which to access IoT data and on which custom IoT applications (or host-acquired IoT applications) can be developed. SaaS can be provided on top of the PaaS solutions to offer the provider's own SaaS platform for specific IoT domains such as smart city, healthcare, video surveillance etc.

It is well understood that cloud computing platforms are well suited for hosting IoT applications as they offer an elastic hardware resources (e.g. CPU, Storage, and Network) that can be scaled on-demand for handling large quantities of data from IoT applications with uncertain volume, variety, velocity, and query types [23, 24]. These two technologies are inherently and increasingly getting entwined with each other. A typical relationship of IoT-based applications and cloud computing is shown in Figure 1. Though, IoT is exciting on its own the real innovation and value from IoT can be harnessed by combining it with cloud computing.

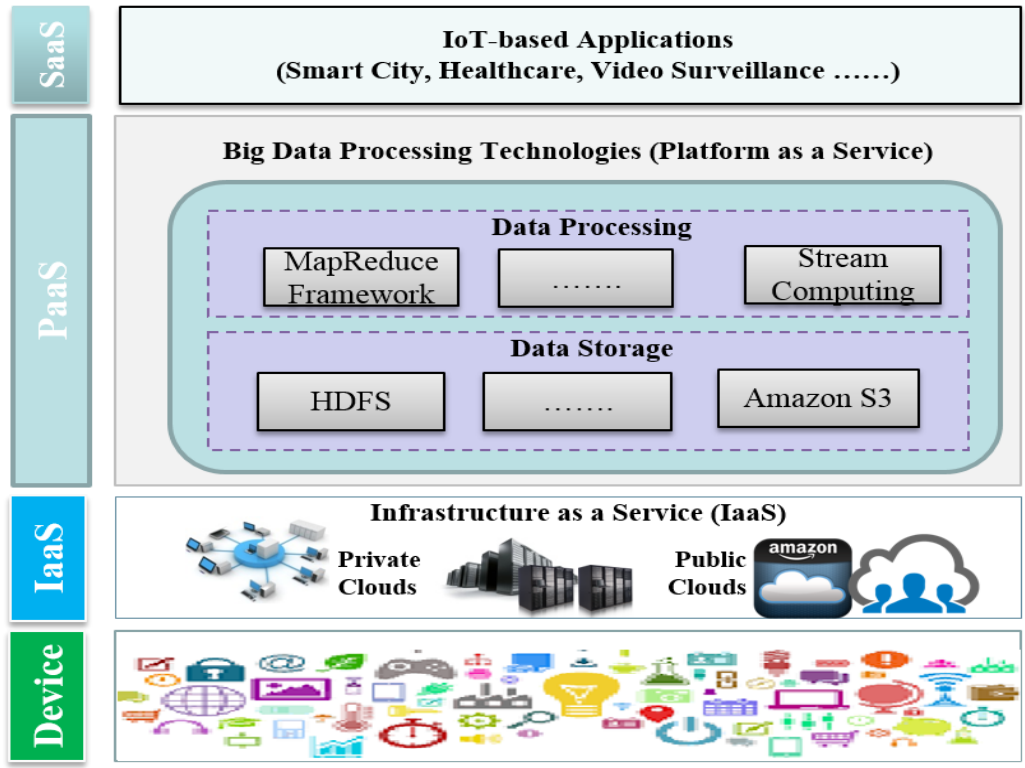

Figure 1.    Relationship of IoT-based application and cloud computing

2.2 Big Data Processing Platforms: MapReduce

The nature of the IoT, or properly speaking, of the data that IoT devices generate, leads to the "big data approach". This is the idea of running data processing in a scale-out fashion on commodity hardware, using distributed data processing framework such as MapReduce for data incentive IoT-based applications.

MapReduce is a predominant framework for large-scale distributed data processing based on the divide and conquer paradigm[9]. MapReduce works by breaking the processing into map and reduce phases. Map task and reduce task are executed in parallel on the different machines within the Hadoop cluster by MapReduce framework. Map performs filtering and sorting operations, and reduce performs summary operations. The user can specify map/reduce functions, and types of input/output.

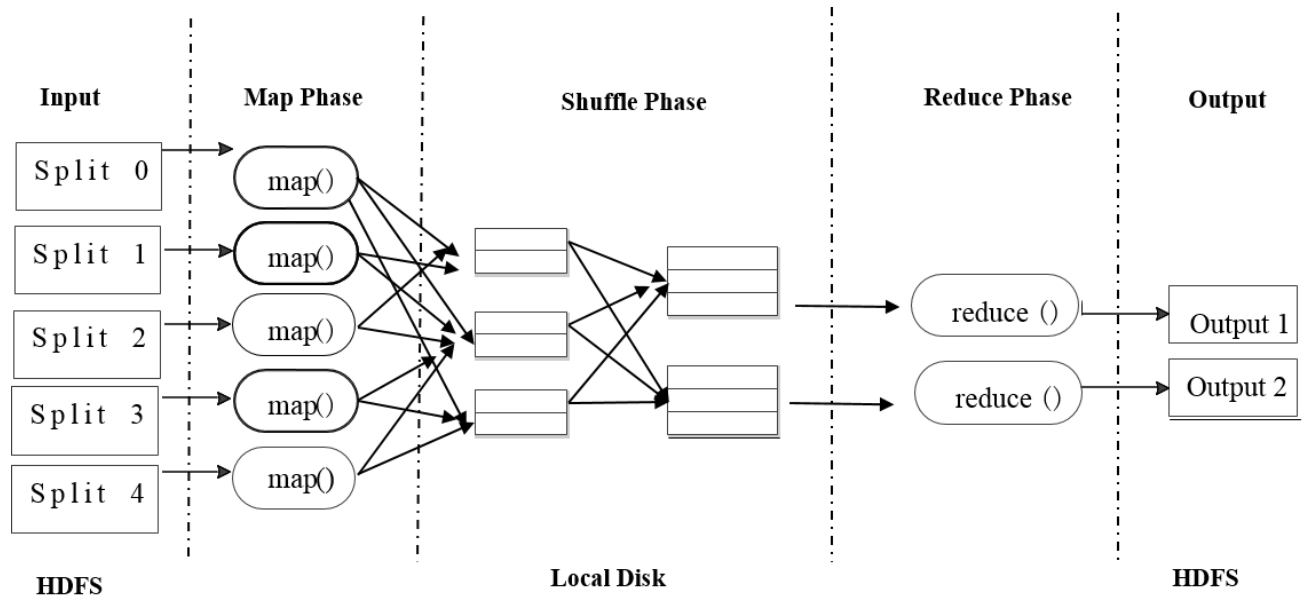

Figure 2. The Structure of MapReduce Model

Figure 2 illustrates the overall process of MapReduce. Input data stored on the Hadoop Distributed File System (HDFS) are split into fixed-size blocks, and each block is allocated to a map. Then user-specified map processes each key-value pair in the block; and outputs the result as a list of key-value pairs. The output of the map is partitioned by the key, and the grouped records are stored in the local disk and transferred to the different reducers, respectively (called shuffle). Then, the transferred records are merged and sorted in the node where a reduce task performs. Each reduce task sequentially reads key-value pairs, and processes them by the user-specified reduce function. Finally, the output records of the reduce task are written to the HDFS.

Figure 3 presents the high level workflow of how a MapReduce job is executed in Hadoop. Specifically, MapReduce processing in Hadoop is handled by the JobTracker and TaskTracker daemons. The JobTracker maintains a view of all available processing resources in the Hadoop cluster and, as application requests come in, it schedules and deploys them to the TaskTracker nodes for execution. As applications are running, the JobTracker receives status updates from the TaskTracker nodes to track their progress. The JobTracker needs to run on a master node in the Hadoop cluster as it coordinates the execution of all the incoming jobs. An instance of the TaskTracker daemon runs on every slave node in the Hadoop cluster, which means that each slave node has a service that ties it to the processing (TaskTracker) and the storage (Data Node), which enables Hadoop to be a distributed system. As a slave process, the TaskTracker receives processing requests from the JobTracker. Its primary responsibility is to track the execution of MapReduce workloads happening locally on its slave node and to send status updates to the JobTracker.

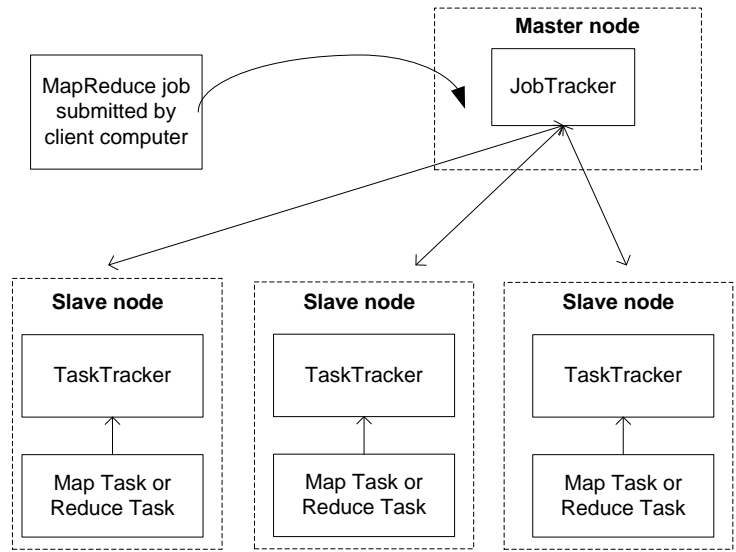

Figure 3. The MapReduce Workflow in Hadoop

2.3 Requirements for Modelling IoT-based Applications

There are two main categories of issues with support for modelling IoT-based applications within a simulator. The first issue is from the perspective of application configuration and modelling, and another is from cloud service providers' perspective. In the following sections, we discuss these issues and provide features that allow support of IoT-based applications for simulation of cloud computing environments.

2.3.1 Application Requirement

As it can be noticed from previous section, IoT-based application generally processes large data sets stored in clouds after being collected from different devices. The simulator should thus allow modelling different IoT-based applications depending on used big data processing platforms such as MapReduce. For example, a MapReduce-compatible IoT application may consist of one or more jobs that will be split into a specific number of chunks of data. These chunks are processed as map tasks at the beginning, and thereafter, the intermediate output of map task is processed by the corresponding reduce task.

2.3.2 Big Data Processing Requirement

To support IoT applications, big data processing technologies plays a central role. Hence, it is mandatorily required for the proposed simulator to meet the big data processing requirement. Depending on various IoT-based applications, the simulator should offer the capability that uses different big data processing technology to support batch processing or stream processing on big data. Also, it should allow modelling and simulating the execution of multiple jobs simultaneously in a scalable manner as it happens in the real world.

2.3.3 Network and Processing Infrastructure Requirements

IoT-based application requires different type of storages that is commonly ambient in cloud-based data centres to store content from the various devices, thus, a storage layer should be modelled to simulate storage (such as Amazon S3, Azure Blob Storage, Hadoop HDFS) and retrieval of any amount of data, subject to the availability of network bandwidth. It is obvious that accessing files in storages at run-time incurs additional delay for IoT-based application execution. This is due to the latencies between the nodes and storages when transferring the data files through the IoT network. Hence, network requirement to model the aforementioned delay is required as well.

3. RELATED WORKS

Over the last decades, many simulation frameworks have been developed to facilitate researches on the behaviour of large-scale distributed systems for hosting various application services (e.g., social networking, web hosting, scientific applications, and content delivery). It is well understood that simulators offer an environment where performance evaluation studies can be conducted in a repeatable and controllable manner.

To the best of our knowledge, there is no simulator available which specifically targets the IoT environment. However there are some closely related in terms of cloud simulator and MapReduce simulator.

3.1 Cloud Simulator

Popular cloud simulators that are capable of simulating and modelling distributed system are typically classified into the following categories:
- Grid Computing: the typical simulation toolkits are GridSim[25], MicroGrid[26], GangSim[27], SimGrid[28], and OptorSim[29].
- Peer-to-Peer network models such as structured and unstructured overlay networks were simulated in PlanetSim[30]. In one study[31], PlanetSim was integrated with GridSim for evaluating the performance of decentralized and coordinated scheduling of scientific applications across multiple computational sites (clusters, supercomputers, etc.).
- Cloud computing model were simulated in GreenCloud[32], iCanCloud[33], Cloudsim[34] and its variants (CloudAnalyst[35], NetworkCloudsim [36], EMUSIM[37], MDCSim[38]) has been described and compared[39].
  - GreenCloud, which is a packet-level simulator (developed by extending NS-2) is capable of modelling behaviours of network links, switches, gateways, and other hardware resources (CPU and storage) in a cloud datacentre. The goal of this simulator is to simplify performance tests of energy-aware scheduling algorithms in cloud environments. GreenCloud is a packet-level simulator hence it requires extra memory and processing power to create and transmit packets across simulation entities.
  - iCanCloud: It is a simulation platform which is oriented towards the simulation of a wide range of Cloud Computing systems and their underlying architectures. It has the ability to model and simulate large environments (thousands of nodes) and distributed applications with a customizable level of detail.
  - Cloudsim is one of the widely used discrete event (its definition is detailed in Section 4.3) simulation frameworks as it is highly extensible and flexible. It provides models for all hardware resources including CPUs (virtual machine), storage and networks (network contention and delays) within multiple datacentres. Cloudsim has extensive support for application (e.g., scientific and web hosting applications) scheduling level simulation, as it provides cloud broker and cloud exchange (for federated datacentre resource pooling) entities.

3.2 MapReduce Simulator

Further to the above cloud simulators, some researches designed and implemented the simulation tools specifically targeted for MapReduce framework. Such MapReduce simulators include:
- MRPerf[40], which can serve as a design tool for MapReduce infrastructure and can help in designing new high performance MapReduce setups, and in optimizing existing ones. However, it can simulate limited behaviours of the Hadoop framework and [41] claimed that accurate results for jobs of different type of algorithms or different cluster configurations cannot be generated based on testing they performed on the MRPerf code.
- Mumak[42]: It is an open source Apache's MapReduce simulator which use data from real experiment to estimate completion time for Map and Reduce tasks with different scheduling algorithms. In cases which data from real experiments do not exist, Mumak cannot estimate completion time for Map and Reduce tasks.
- SimMR[43]: was developed in HP lab. It can replay execution traces of real workloads collected in Hadoop clusters (as well as synthetic traces based on statistical properties of workloads) for evaluating different resource allocation and scheduling ideas in MapReduce environments.
- MRSim[41]: It is a discrete event based MapReduce simulator. It is able to simulate different type of MapReduce applications with the ability to study with good accuracy the effect of dozens of job configuration parameters on the job performance. However, it was modelled and simulated using SimJava discrete event engine that has intrinsic weakness such as increased kernel complexity [44] and lack of support of some advanced operations[34]. Because of this, the SimJava layer has been removed from Cloudsim 2.0 onwards.
- MR-Cloudsim[45] was developed (by extending Cloudsim) for simulating MapReduce big data processing model. However, MR-Cloudsim has several limitations as it only supports simplistic, single-state Map and Reduce computation. Further, it lacks support for network link modelling, which is a critical element affecting the performance of MapReduce applications. Also, there is lack of support for allowing multiple MapReduce applications.

Although, the aforementioned two types of simulators were widely adopted in the study of the behaviours of cloud computing and MapReduce in distributed computing environment, they obviously lack the support of modelling and simulation for IoT applications. Hence our IOTSim will focus on simulating IoT environment and aim to offer the following advantages compared to those existing simulators.
- support for simulation of IoT big data processing using MapReduce model or steam model in cloud computing environment
- support for modelling and simulation of large scale multiple IoT applications to run simultaneously in a shared cloud data centres
- support for modelling network and storage delays existing in the processing of IoT applications

4. DESIGN AND IMPLEMENTATION OF IOTSIM

Cloudsim [34], is an extensible simulation toolkit that enables modelling and simulation of cloud computing environments and application provisioning. It has many features, which make us choose it for building our simulator for analysing Iota Application. To be specific, Cloudsim supports modelling and creation of one or more virtual machines (VMs) on a simulated node of a datacentre with different hardware configurations, cloud-based tasks and their mapping to suitable VMs. It also allows simulation of multiple datacentres to enable a study on federated and associated policies for migration of VMs for reliability and automatic scaling of applications. In addition, Cloudsim helps in modelling user applications having independent jobs, and design and analysis of different hardware configurations, VM provisioning and scheduling policies. Hence, Cloudsim can pave the way for us to design and implement our simulation tool specific for IoT-based applications. In fact, Bashar[15] had done a critical evaluation on various cloud computing simulators and his study has concluded that Cloudsim is the best choice if research has to be done by using a simulation software. In the following section, we will details how we design and implement IOTSim by extending Cloudsim.

4.1 Proposed Architecture

Illuminated by the works of Cloudsim[34], our IOTSim simulator is designed using the layered architecture with support for big data processing framework. Figure 4 shows components of the Cloudsim architecture with the key elements of IOTSim (shown by dark boxes). In this section, we outline the general layered architecture of IOTSim. The detailed design and functionality of our proposed IOTSim's components will be discussed in the later sections.

- Cloudsim Core Simulation Engine Layer: the bottommost layer, which is a simulation engine that supports several core functionalities, such as queuing and processing of events, creation of Cloud system entities (services, host, datacentre, broker, and virtual machines), communication between components, and management of the simulation clock.
- Cloudsim Simulation Layer: this layer provides support for modelling and simulation of virtualized Cloud-based datacentre environments including dedicated management interfaces for virtual machines (VMs), memory, storage, and bandwidth. The fundamental issues such as provisioning of hosts to VMs, managing application execution, and monitoring dynamic system state are handled by this layer. This layer consists of several sublayers that model the core elements of Cloud Computing. The bottommost sublayers model datacentre, cloud coordinator, and network topology. These components help in designing Infrastructure-as-a-Service (IaaS) environments. The VM Services and Cloud Services provide the functionality to design resource (VM) and management and application scheduling algorithms.
- Storage Layer: this layer supports modelling different type of storage such as Amazon S3, Azure Blob Storage, and HDFS etc. where large datasets generated from devices are stored. In run time, IoT-based applications copy the data files from these storages, and write the intermediate data files to these storages when need. A storage delay will be incurred at this layer.
- Big Data Processing Layer: it includes two sub-layers. MapReduce sublayer is to support applications where a batch-oriented data processing paradigm is required while Streaming Computing sublayer aims to support applications that need a real-time processing paradigm. Depending on which IoT-based applications the customer will use, it can support processing for the big data generated from IoT devices or sensors using MapReduce or streaming computing model. Due to the limitation of Cloudsim mentioned in Section 3, the Big Data Processing Layer has been highly demanded in order to simulate and analyse IoT-based Applications. Take the MapReduce-compatible applications as an example, a MapReduce model need be fully implemented here where a set of new classes or entities such as JobTracker, TaskTracker, Mapper and Reducer work as does the real Hadoop and a series of events occur in some specific order to finish a Map/Reduce process. This layer plays an integral role in support big data processing towards IoT-based applications.
- User Code Layer: the top-most layer which exposes basic entities for hosts (number of machines, their specification and so on), IoT-based applications configurations (Job Length and their requirements), VMs, number of users and their application types, and broker scheduling policies. This layer helps users to define their own simulation scenarios and configurations for validating their algorithms.

4.2 Design Considerations

4.2.1 Application Model

The application models in IoT environment can vary from building and home automation to wearables applications areas. The current generation of IoT applications (such as smart city, smart healthcare, and video surveillance) combines multiple independent data analytics models, historical data repositories, and real-time data streams that are likely to be available across geographically distributed datacentres (both private and public). Typically, such applications need to process large amounts of data by using parallel big data processing technologies such as MapReduce.

Most simulators for cloud computing generally offer limited support for modelling execution of parallel and distributed applications. For example, Cloudsim allows modelling a task by using a programming structure called "Cloudlet", which only represents single and atomic computation needs. This approach is obviously not appropriate for IoT applications scenarios. In order to support IoT applications simulation, one approach that could be used is to introduce two types of Cloudlet, one is MapCloudlet, and the other is ReduceCloudlet. The details of the two types of Cloudlets are described in the next Section 4.3.1. Both of them inherits from Cloudsim Cloudlet and has their own specific attributes and will be paired during the simulation. Also, ReduceCloudlet always runs after MapCloudlet of the same input.

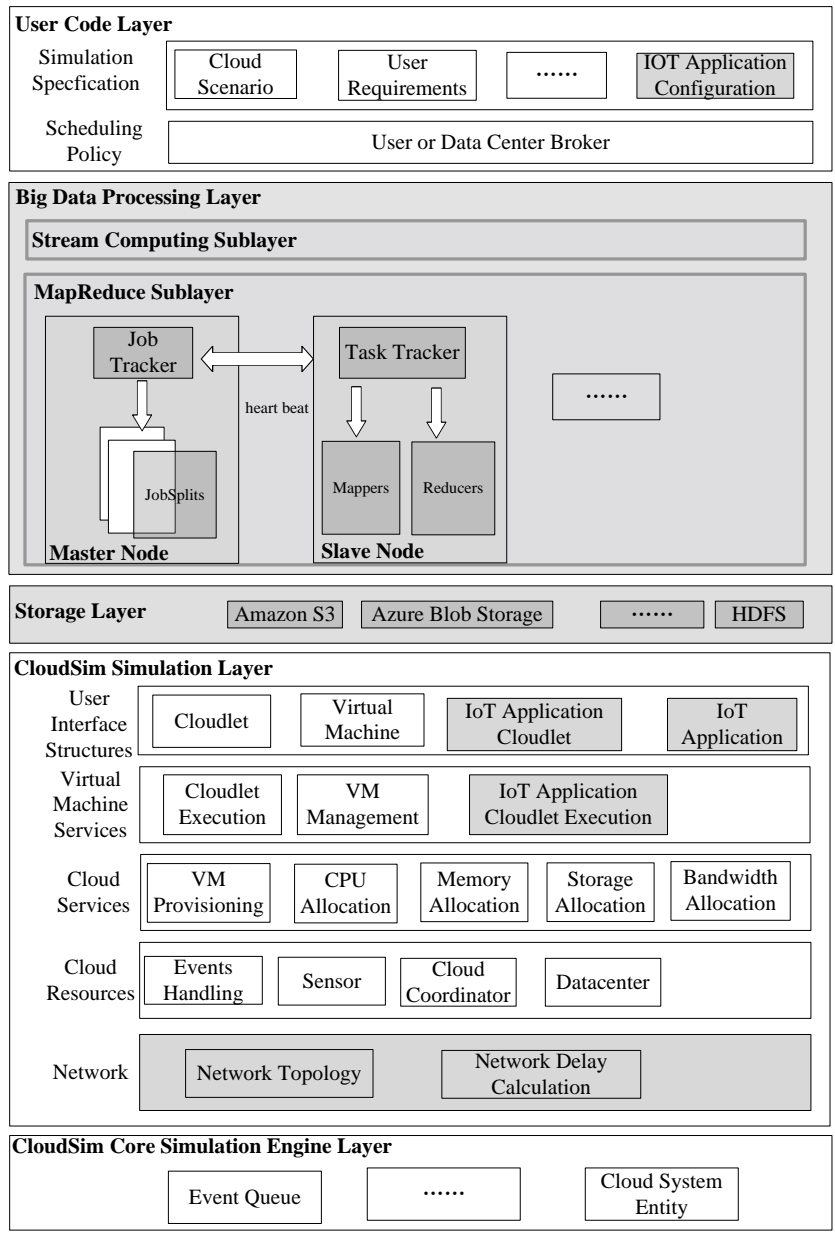

Figure 4.    The Proposed Layered Architecture of IOTSim

### 4.2.2 Big Data Processing Model

When dealing with IoT-based applications, parallel big data processing system has become the key to IoT applications. MapReduce, as a predominant big data processing framework is largely utilized by IOT applications. In Cloudsim, when Cloudlets come in, they are simply submitted to the cloud datacentre via broker. However, it doesn't suit for submitting MapReduce job in the same way due to the complexity of MapReduce workflow. Since JobTracker and TaskTracker play an integral part during the MapReduce processing as depicted in the above section, the IOTSim simulator needs to fully model and implement them as well as other notable features that MapReduce model has. Such features include: 1) there will be lots of communication and interaction between them on top of Cloudsim during the execution; 2) every separated Map task output has one Reduce operation; 3) Reduce operation has to come after Map operation of corresponding input; 4) multiple MapReduce jobs can be submitted and run simultaneously in a shared cloud-based big datacentres.

### 4.2.3 Network and Storage Model

As mentioned, a large amount data that are generated from devices or sensors are stored in the Storage Layer. In runtime environment, a Map instance (mapper) operates in each slave node. It copies data which is saved in the above Storage Layer to its own local hard disk. When the data is copied and saved in the local hard disk, mapper starts processing the allocated map task by TaskTracker. Thereafter, the intermediate output will be generated and are associated with the Reduce instance. The paired reducer reads the intermediate output and start processing the allocated reduce task. The final output will be written into the Storage Layer. Therefore, there are two typical network delay incurred that affect the performance of map or reduce task. In this scenario, network and storage model must be represented in a IOTSim simulator. One of feasible methods is to calculate the network consumption when copying data from Storage Layer by mapper or copying intermediate output from local disk by reducer. It is enough to guarantee the accuracy of simulator when calculating and presenting the network delay incurred during the processing.

### 4.3 Design of Entities/Classes

We would like to introduce the core components of Cloudsim before we present the design of IOTSim. There are two main parts in Cloudsim simulation: entity and event:

- Entity: refers to something which can individually and independently exists. It is able to send messages to other entities and process received messages as well as trigger and handle events. Each entity is initiated at the beginning and shutdown at the end during the simulation.
- Event: represents a simulation event which is passed between the entities in the simulation. Each event carries all the related information about an event between two or more entities such as event type, init time, time at which the event should occur, finish time, time at which the event should be delivered to its destination entity, the source entity and the destination entity as well as the data that has to be passed to the destination entity. Since Cloudsim is a discrete event-driven simulator, it is dependent of the series of event that occur in some specific order. Without this order the simulation is impossible.

#### 4.3.1 Application Modelling Design

To model the MapReduce for IoT-based applications, the following classes have been designed.
- MapCloudlet: This class, which is inherited from Cloudlet, models the atomic map task which will be submitted by DatacenterBroker and executed in VM. It extends Cloudlet with specific attributes.

- ReduceCloudlet: This class, which is inherited from Cloudlet, models the atomic reduce task which will be submitted by DatacenterBroker and executed in VM. It extends Cloudlet with specific attributes.

### 4.3.2 MapReduce Modelling Design

To model a MapReduce process and behaviour within a datacenter, the following classes/entities have been added to the IOTSim.

- JobTracker: represents an entity which receives the jobs summited by user, gets the data from storage, splits the jobs according to user requirements and schedules them to the TaskTracker node for execution. As applications are running in VM, the JobTracker receives status updates from the TaskTracker nodes to track their progress. If JobTracker finds all Mappers finish their tasks successfully, then it will communicates with TaskTracker to launch the corresponding Reducer to execute reduce task. It produces the intermediate output results after each Mapper finishes its work and provides input for Reducer
- TaskTracker: These classes represent entities which receive processing requests from the JobTracker. Its primary responsibility is to track the execution of map and reduce tasks happening locally on its slave node and to report the status updates of each map and reduce tasks to the JobTracker. TaskTracker manages the processing resources on each slave node in the form of processing slots — the slots defined for map tasks and reduce tasks. It schedule the split map tasks and reduce tasks to the corresponding Mapper and Reducer, and monitors the status updates of them
- Mapper: These classes represent entities which receive the request of TaskTracker and communicate with IoTDatacenterBroker to submit the corresponding map tasks to be executed in datacentre. It frequently reports the status of map tasks to TaskTracker
- Reducer: These classes represent entities which receive the request of TaskTracker and communicate with IoTDatacenterBroker to submit the corresponding reduce tasks to be executed in datacentre. It frequently reports the status of reduce tasks to TaskTracker.

Figure 5 shows the above main classes or entities and their interrelationship in IOTSim.

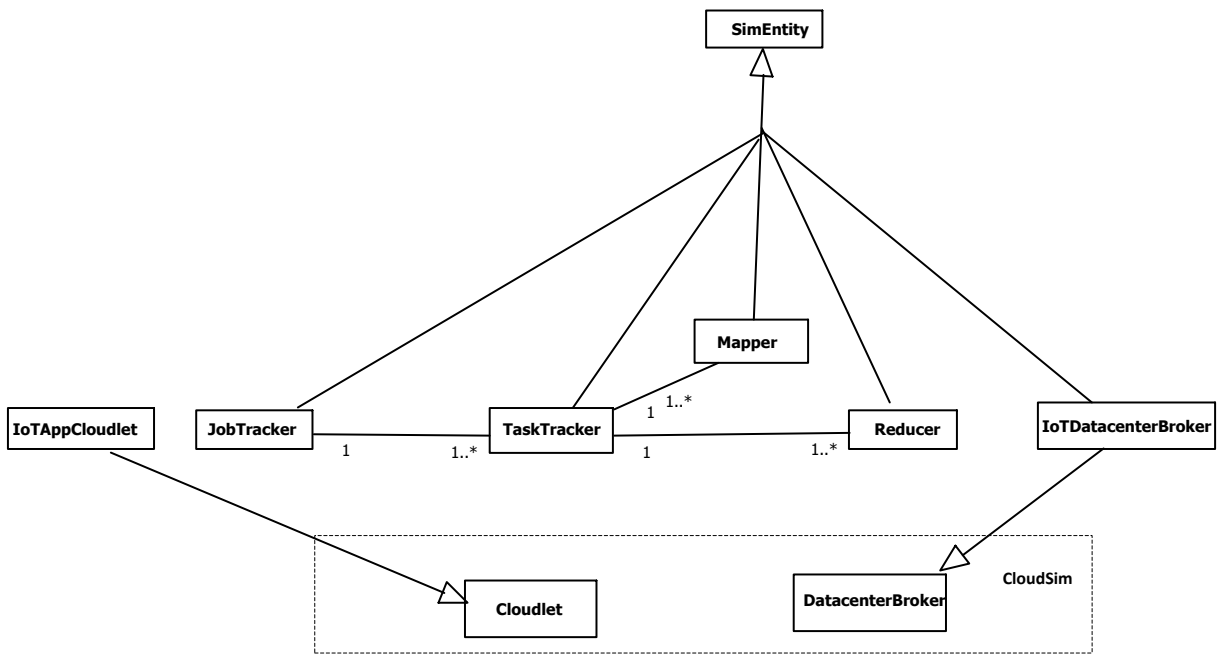

Figure 5. Class diagram of IOTSim

Figure 6 summaries the workflows of these entities. Once an IoT-based job has been submitted, JobTracker begins to track a simulated job and split the job. It creates one map task for each split. Then, TaskTracker starts to track each map task and sends messages to the JobTracker via event mechanism which reports the task status to the JobTracker. Once TaskTracker is aware that the current tasks are finished, then, it starts to run a new task. If all the map tasks are finished, then TaskTracker report it to the JobTracker, and is ready to run the corresponding reduce task.

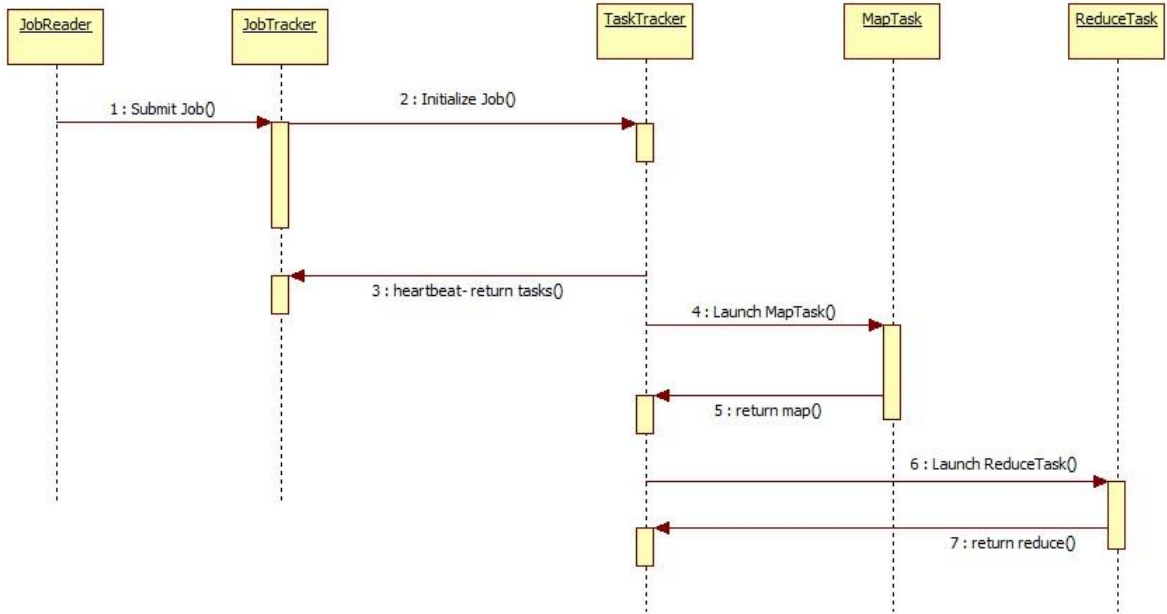

Figure 6. Sequency diagram: communication between IOTSim entities

4.4 Implementation

As mentioned before, our simulator extends Cloudsim. Hence, the implementation consists of two parts according to the aforementioned design decisions: modification and addition. Modifications are done on the original Cloudsim code including Datacentre, CloudTag, and DatacenterBroker etc. Upon the requirements stated in Section 2, the big data processing paradigm need to be implemented.

Figure 7 shows the flow of how IOTSim works. Firstly, IOTSim initializes a series of entities, for example, Datacentre, Broker, JobTracker and TaskTracker. A specific number of VM has been created with the pre-defined configuration (stated in Table II). It also accepts multiple user-defined MapReduce jobs. When MapReduce jobs come in, JobTracker split them into a number of blocks and schedule them to TaskTracker for execution. TaskTracker allocates the split job into the corresponding Mapper and Reducer. Then, Mapper and Reducer submit the corresponding map task and reduce tasks to the VMs where they are executed. The status of the map task and reduce task are reported to TaskTracker. In the runtime, the enhanced DatacenterBroker object works linking each object. When the linking ends, the simulator takes an action on captured event time. An event occurs when Cloudsim creates, executes, and terminates each object such as datacenter, broker, VM, JobTracker and TaskTracker. runClockTick() function works checking each SimEntity object and this state which is runnable at the event time. If the state is runnable, each SimEntity object classifies its own operable events. Each entity checks simulating tag and operates each request. Each object has one of various tags. They consist of entity creation, acknowledge, characteristic setting, event pause, move, submit, migration, termination and etc. At the event time that the cloudlet process is submitted, the simulator calculates all submitted cloudlet's processing time. During the event processing, a new event may be created. When new event is created, send() or sendNow() function is called. These functions notify that one event time is created. When the all event time is over, then simulating ends and the simulation result is reported. The simulation reports consist of each MapReduce jobs' information (name, split number, job length etc.), status, executed in which datacentre and which VM, and processing result (job id, VM id, start time, execution time, finish time, VM computation cost etc.).

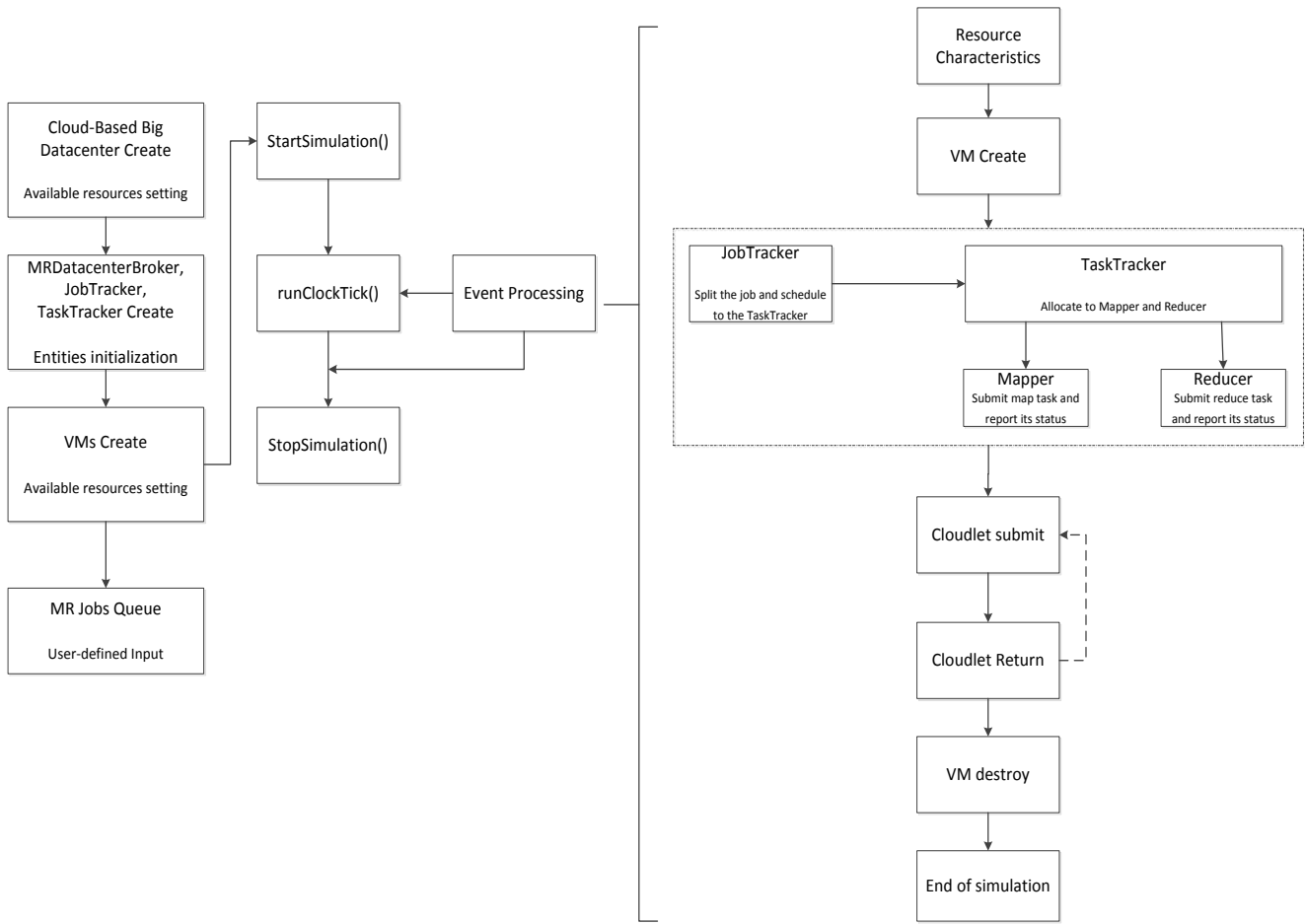

Figure 7.    The Basic Workflow of how IOTSim works

## 4.5  Extensions to Cloudsim

We have extended and enhanced Cloudsim with new functionality so that it can support executing multiple CloudletLists sequentially. The work originated from a functionality limitation existing in current Cloudsim. Internally, the broker has a very simple operation and has a single CloudletList, which means if you submitted multiple CloudletLists to the broker, they are always merged to a single list, and they are handled as if only one submission were made. However, it is not suitable for a real MapReduce framework as the reduce operation can only start after the corresponding map operation.

In order to solve this problem, we implemented a new broker called IOTSimBroker which inherits the original broker in Cloudsim. This new broker can accept the multiple Cloudlets and execute them sequentially. By virtue of this, the new broker can guarantee reduce task starting before corresponding map task has ended.

5. EVALUATION

Currently, we implemented MapReduce model as one of the big data processing paradigm while keeping the stream computing model as future work. To evaluate the efficacy of IOTSim, we conduct a number of experiments and collect results. The experiment results were collected in a machine that had one Intel i7-5500U Core 2.40GHz and 16GB of RAM memory. All of these hardware resources were made available to a VM running Windows 7 SP1 that was used for running the simulator.

5.1 Experiment Scenario

IoT applications are enabling smart city initiatives all over the world. Smart city includes many different components (i.e., smart transportation, smart healthcare, smart energy)[46], where big data processing technologies such as MapReduce play an integral part. An example scenario could be from smart transportation that a city council is planning to optimise and expand its road network. For such a scenario to be feasible, it is important that large datasets such as road network, road traffic, commuter requirements etc. are collected and stored in the cloud infrastructure. Further, the council needs to process and analyse this big data stemming from IoT devices to extract relevant knowledge such as highly utilized roads, traffic patterns, risky roads etc. Using such datasets is not an easy task due to its huge size and non-standard format [47], hence, the MapReduce paradigm is used to speed up data extraction, indexing, and querying from this big data in such scenarios. For our experimental evaluations, we have considered the smart city application described earlier for modelling. We considered there is one such application job or multiple jobs coming as an input into our simulated environment. The incoming jobs are processed using MapReduce approach which produces the corresponding output once the jobs are finished. The output presents the related information regarding the jobs including job type, job length, job size, the VM ID where map task or reduce task is being processed, start time, execution time, and finish time etc.

For comparisons, we have summarized four group experiments. In each group, we change an independent variable, while keeping others constant. Hence, we can clearly observe the result and make an appropriate analysis on how dependent variable is impacted by independent variable in details and if it matches the real world.

In the experiment, we define a series of key independent variables and dependent variables. independent variable include datacentre configuration, VM configuration, VM number, job configuration, MR combination while dependent variable include average execution time, maximum execution time, minimum execution time, make span, delay time, VM computation cost, network cost. These independent variables are the main impact factors that affect the above dependent variables. They will be detailed in the next section.

It is worth noting that all of the above group experiments (which are detailed in Section 5.4) are with two cases considered.

- Without Network Delay: in this scenario, JobTracker splits the MapReduce jobs according to user requirements and schedules them to the TaskTracker node for execution immediately when it receives the jobs summited by user. Also, when all the Mappers finish the map task successfully, the corresponding reduce tasks begin to execute immediately. This scenario means no network delay is incurred during the whole period that the job is running in the simulated big data environment.

- Network Delay: in this scenario, JobTracker firstly gets the data from storage (HDFS) for each MapReduce job when the simulation begins, this causes the first delay that job starts after the simulation clock. By subtracting the start time of map task and the start time of simulation clock time (also refer to the formula of Delay Time stated in Section 5.3.5), the result can be visualised. When all Mappers finish the map tasks, each Mapper will produces an intermediate output, then, the corresponding Reducer begins to work after it reads the intermediate output (in Hadoop, this is the shuffle process). Obviously, this causes the second delay. The result can also be visualised by subtracting the start time of reduce task and the finish time of the corresponding map task.

5.2 Independent Variables

5.2.1 Datacenter Configuration

Datacentre models the physical hardware that is offered by big data application provider. It encapsulates a set of computing hosts that can either be homogeneous or heterogeneous with their hardware configurations (memory, CPU, storage) where specific number of hosts and VMs are generated and run. Table Ⅰ lists the datacentre configuration that is going to be consistent throughout the entire evaluation.

TABLE I. DATACENTER CONFIGURATION

| pesNumber | 500 |
|---|---|
| Ram | 20480 |
| Storage (MB) | 1000000 |
| Bandwidth | 1000 |
| MIPS | 1000 |

In Cloudsim, there are a few other parameters including OS, system architect, and VMM to be defined for initialization. They are not factors that can affect the aforementioned dependent variables. Hence, such parameters are irrelevant in our experiments.

5.2.2 VM Configuration

VM component stores the following characteristics related to a VM: processor, memory, storage size, bandwidth and MIPS in Cloudsim. It should be submitted to the broker ahead of the simulation. VM parameters are monitored in Cloudsim, which means each sum of VM parameter must be less than the corresponding datacentre configuration. For a simplified reason, we define three types of VM (Small, Medium, and Large) which is compatible with typical computing infrastructure in Amazon etc. provider. The configuration of three types of VM is listed in Table Ⅱ.

TABLE II. VM CONFIGURATION

| VM Type | Small | Medium | Large |
|---|---|---|---|
| Image Size (MB) | 10000 | 20000 | 40000 |
| Ram | 512 | 1024 | 2048 |
| MIPS | 250 | 500 | 1000 |
| Bandwidth | 1000 | 1000 | 1000 |
| pesNumber | 1 | 2 | 4 |
| VMCostPerSec ($) | 1 | 2 | 4 |

### 5.2.3 VM Number

Further to the above VM Configuration itself, we will also specify VM Number. This independent variable refers to the quantity of VM in Datacentre. VM, which is hosted in the host, is the basic computing unit where the jobs will be really processed. In our experiment, we will change the VM Number when required. For the sake of simplicity, we set the VM Number to be 3, 6 or 9. We can definitely set a very large number of VM, however, it is limited by the capacity of the Datacentre.

### 5.2.4 Job Configuration

In Cloudsim, the complexity of an application is abstracted in terms of its computation requirements. Every application service has been modelled by Cloudlet. It has a pre-assigned instruction length and data transfer overhead that it needs to undertake during its life cycle. For the sake of simplicity, we define three types of job configuration as following.

TABLE III.    JOB CONFIGURATION

| Job Type | Small | Medium | Big |
| :---: | :---: | :---: | :---: |
| Job Length (MI) | 362880 | 725760 | 1451520 |
| Data Size (MB) | 200000 | 400000 | 800000 |

### 5.2.5 MR Combination

This represents the specific joining number of map task and reduce task. For the proposed IOTSim, when datacentre receives a job with Job Length (MI) and data size specified, the joining number of map task and reduce task in each job (which we call MR Combination) should also be considered throughout the experiment. For example, M1R1 means there are one map task and one reduce task for this job. In the same way, M40R1 means there are 20 map tasks and one reduce task for this job.

## 5.3 Dependent Variables

### 5.3.1 Average Execution Time

It refers to the average execution time of Map/Reduce job. Its value is given by

$$\text{Average Execution Time} = \frac{\sum_{i=1}^{nm} et_m(i)}{nm} + \frac{\sum_{j=1}^{nr} et_r(j)}{nr}$$

, where $et_m(i)$ is the execution time of map task $i$ and $et_r(j)$ is the execution time of reduce task $j$. $nm$ means the number of map tasks in this job, and $nr$ represents the number of reduce tasks in this job.

### 5.3.2 Maximum Execution Time

It means the maximum execution time of MapReduce Job. Its value is given by

$$\text{Maximum Execution Time} = \max(et_m(i)) + \max(et_r(j))$$

### 5.3.3 Minimum Execution Time

It represents the minimum execution time of Map/Reduce Job. Its value is given by

$$\text{Minimum Execution Time} = \min(et_m(i)) + \min(et_r(j))$$

### 5.3.4 Make Span

It is the time span of Map/Reduce job from start to finish. It is calculated by

$$\text{Make Span} = ft_r(nr)$$

, where *ftr(nr)* means the finished time of reduce task *nr*.

### 5.3.5 Delay Time

It means the discrepancy between reduce task starts and map task starts. It is calculated by

$$\text{Delay Time} = st_m(nm) + st_r(nr) - ft_m(nm)$$

, where *stm(nm)* means the start time of map task nm and *str(nr)* means the start time of reduce task *nr*.

### 5.3.6 VM Computation Cost

It refers to the CPU computation cost incurred when VM runs a Cloudlet. It is calculated by

$$\text{VM Computing Cost} = (\sum_{i=1}^{nvm} et_m(i) + \sum_{j=1}^{nvm} et_r(j)) \times VMCost / Unit\ Time$$

, where *etm(i)* means the execution time of VM *i* when running the map task, while *etr(j)* means the execution time of VM *j* when running the reduce task.

### 5.3.7 Network Cost

This dependent variable means the network cost incurred when job task gets data from storage and reduce task gets data from intermediate output of map task. It is calculated by

$$\text{Network Cost} = DelayTime \times NetworkCostPerUnit$$

These dependent variables are very important factors when analysing IoT-based applications. It is not hard to understand they are functions of the aforementioned independent variables, which means the value of these dependent variables change as the independent variables vary. For example, given the same IoT application job, if we increase the VM number within the capability that the datacentre can offer, then, the make span, average execution time etc. may changes because more computing resources are leveraged by map or reduce tasks in the big data processing process.

### 5.4 Experiment Result

In Group 1 experiment, we set specific job configuration (as presented in Table III), VM configuration and VM number, for example job type (Small Job), VM type (Small VM), VM number = 3 and give different MR Combination from M1R1 to M20R1. We then start simulation in either Without Network Delay or Network Delay case respectively. After the simulation successfully finishes, we collect the data of Average Execution Time, Max Execution Time, Min Execution Time, Make Span and VM Computing Cost and Network Cost (only applicable for Network Delay case). The results of Group 1 experiment are presented in the following line chart.

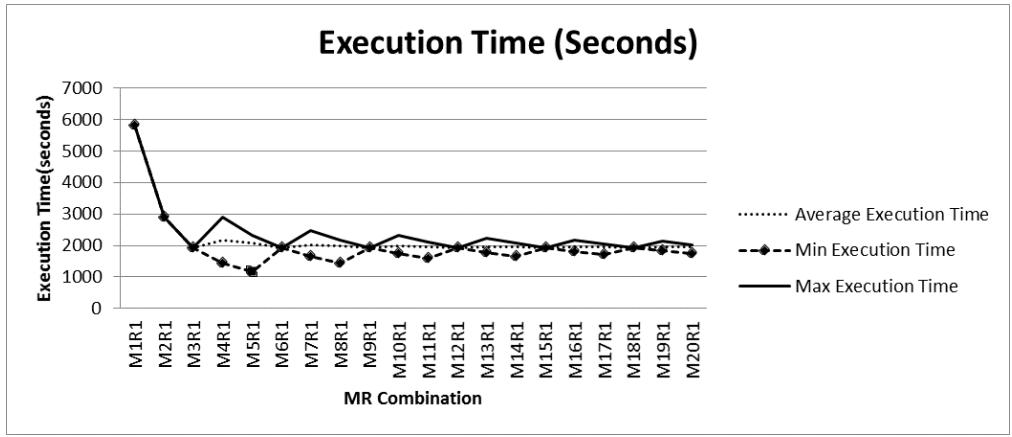

(a)

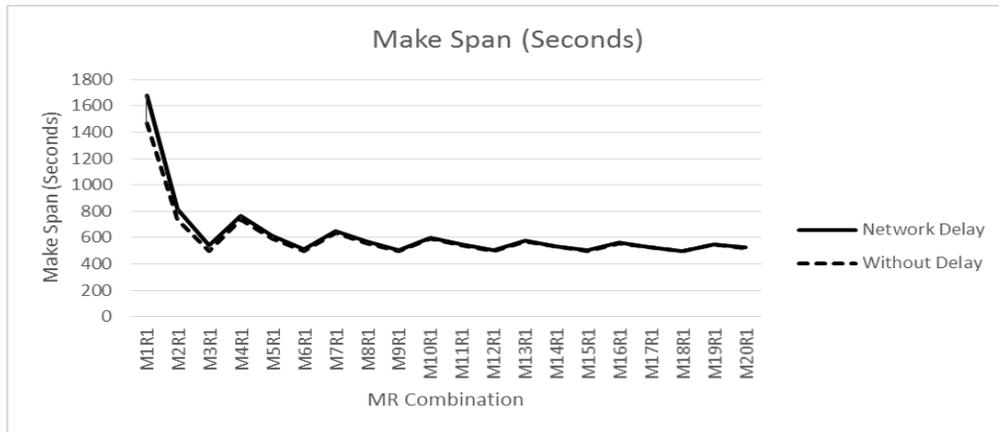

(b)

Figure 8.  Effects of simulation parameters on simulation variables by Changing MR Combination: (a) execution time (Average, Max, Min); (b) Make Span

In Figure 8 (a), generally speaking, Execution Time (Average, Max, Min) fluctuates as MR Combination increases. When number of map task is smaller than the VM number, it can be observed the execution time (Average, Max and Min) is identical because datacentre provides more VMs than the jobs need, which means some VMs are idle. Meanwhile, they decease rapidly because more map tasks are produced and executed simultaneously in datacentre, such that less execution time is consumed. However, if the number of map task is greater than the VM number, execution time (Average, Max, Min) begin to flatten and the discrepancy between them is becoming narrow as MR combination increases. This is because VMs are constrained in computing resources and many map tasks compete to acquire them, such that the effect on reducing execution time by increasing MR combination is getting more and more insignificant.

Figure 8 (b) compares the make span in Network Delay case with that in Without Network Delay case. Clearly, it is observed that the make span of former is slightly larger than the make span of the latter and the disparity between them is getting narrow as MR combination increases. This is because the delay when copying data from storage and getting intermediate data generated by map task incurs in the former case, and the delay becomes less and less as MR Combination increases.

In Group 2 experiment, we set job configuration (Small Job), VM configuration (Small VM) and MR combination (from M1R1 to M20R1) and keep them invariant in the experiment. But this time, VM number varies. In order to get better visualization result, we set the VM number with appropriate discrepancy (i.e., VM number equals 3, 6, 9).

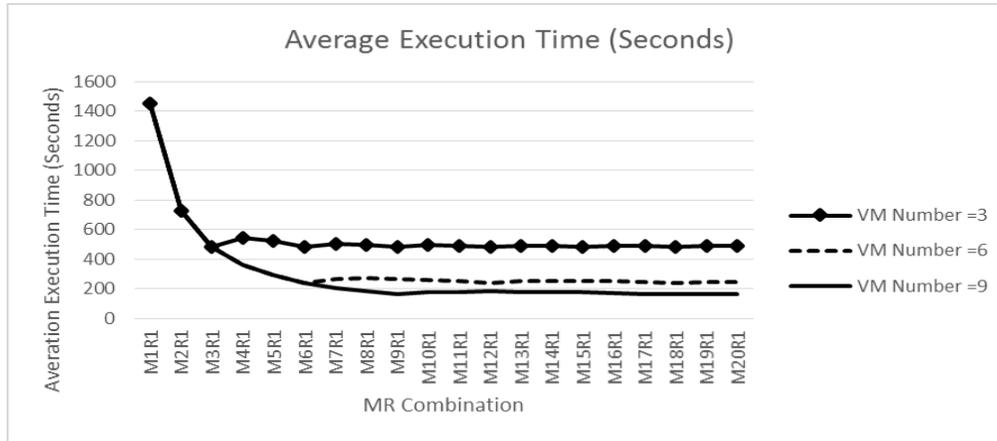

Figure 9. Comparison of Average Execution Time when changing VM Number

Figure 9 shows the comparison of average execution time between these three cases (i.e., VM number = 3, 6, 9). When number of map task is smaller than VM number, their average execution time are equal. Afterwards, the chart shows more VMs results in obvious less average execution time. To be precise, when VM number increases from 3 to 6, the average execution time is reduced by 40% on average and 50% if it further increases to 9. This is because more VM resources are available for processing the same job.

The comparison of network cost is presented Table IV. Interesting, even though VM number has been changed, the network cost is identical. This is because, given the same job, the data size is identical, which results in the same network delay.

| MR Combination | VM Number =3 | VM Number =6 | VM Number =9 |
| --- | --- | --- | --- |
| M1R1 | 2125 | 2125 | 2125 |
| M2R1 | 1416.667 | 1416.667 | 1416.667 |
| M3R1 | 1062.5 | 1062.5 | 1062.5 |
| M4R1 | 850 | 850 | 850 |
| M5R1 | 708.333 | 708.333 | 708.333 |
| M6R1 | 607.143 | 607.143 | 607.143 |
| M7R1 | 531.25 | 531.25 | 531.25 |
| M8R1 | 472.222 | 472.222 | 472.222 |
| M9R1 | 425 | 425 | 425 |
| M10R1 | 386.364 | 386.364 | 386.364 |
| M11R1 | 354.167 | 354.167 | 354.167 |
| M12R1 | 326.923 | 326.923 | 326.923 |
| M13R1 | 303.571 | 303.571 | 303.571 |
| M14R1 | 283.333 | 283.333 | 283.333 |
| M15R1 | 265.625 | 265.625 | 265.625 |
| M16R1 | 250 | 250 | 250 |
| M17R1 | 236.111 | 236.111 | 236.111 |
| M18R1 | 223.684 | 223.684 | 223.684 |
| M19R1 | 212.5 | 212.5 | 212.5 |
| M20R1 | 202.381 | 202.381 | 202.381 |

TABLE IV. COMPARISON OF NETWORK COST WHEN VM NUMBER CHANGES

In Group 3 experiment, we set job configuration (Small Job) and VM number (equals 3) and provide different VM configuration (from Small VM to Large VM) in our experiment. The results of Group 3 experiment are presented in the following figure.

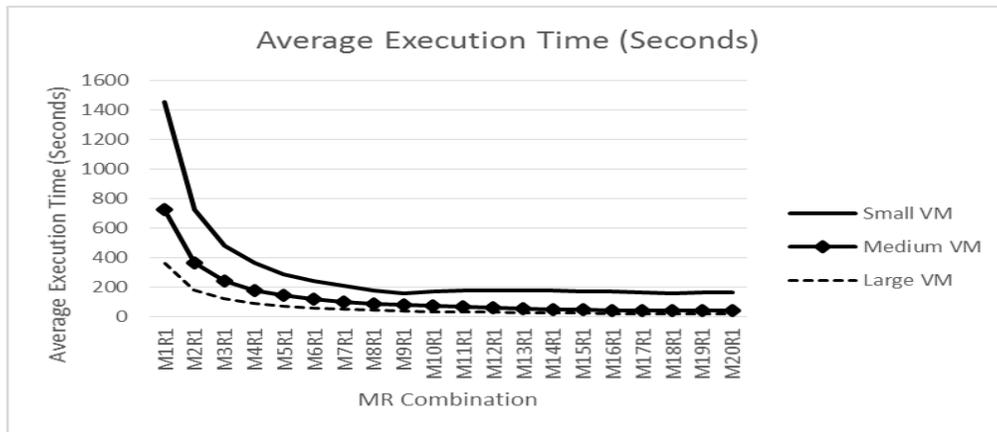

Figure 10. Comparison of Average Execution Time when changing VM Configuratoin

Figure 10 shows average execution time decreases exponentially if we provide higher-profile VM. Precisely, Medium VM gets approximately 60% less average execution time, while Large VM consumes about 80% less average execution time when compared with Small VM. As presented in Table II, Large VM has four times MIPS as much as Small VM, and Medium VM offers twice MIPS than Small VM, hence, higher configured VM can definitely provide more computing capacity.

In Group 4 experiment, VM configuration (Small VM) and VM number (equals 3) are set, while job type varies (from Small Job to Big Job) in our experiment. The VM computation cost is compared in the following figure.

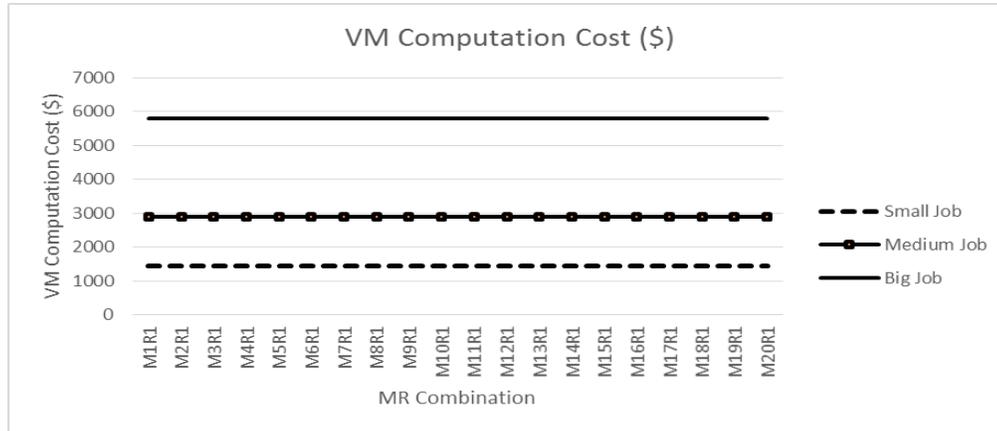

Figure 11.  Comparion of VM Computation Cost when changing Job Configuration

Figure 11 shows Big Job costs VM computing resource twice as much as Medium Job and the VM computation cost of Medium Job is twice than Small Job. As presented in Table III, Big Job doubles its job length (MI) than Medium Job while Medium Job double its job length (MI) than Small Job, hence when the same quantity of VMs with identical VM configuration are offered, higher-workload Job has linearly increased its VM computation cost, which matches the real world.

6. CONCLUSION

Nowadays, increasing IoT-based applications rely on Clouds to run big data processing platform to process massive amounts of data generated from billions of devices and conduct data analytics for knowledge extraction. However, setting-up such environment is very challenging and a tedious task and is expensive in terms of cost and time. To address this problem, we proposed designed and implemented IOTSim. IOTSim allows simulation of IoT application by inherently supporting big data processing system such as with MapReduce to facilitate researchers and commercial organizations to understand and analyse the impact and performance of IoT-based applications. Our simulator is built on top of a widely used simulator, i.e., Cloudsim. However, we have extensively extended and improved the existing functions of Cloudsim.

We performed a number of experiments to evaluate and validate IOTSim based on one of IoT application scenario. IOTSim is able to support simulation of IoT-based big data processing using MapReduce model with the ability to study the correctness and effect of independent variables i.e., VM configuration, VM number, job configuration, MR combination) on the dependent variables (i.e., average execution time, maximum execution time, minimum execution time, make span, VM computation cost and network cost). IOTSim enables the researchers to analyse how a MapReduce-compatible IoT application performs in certain environment. The simulation result provides better perspective to analyse IoT-based applications using MapReduce model in Cloud Computing environment with less cost and time.

In the future, we would like to design and implement stream computing model as a supplementary to big data processing layer and study the service level agreement for IoT-based applications.